\documentclass[traditabstract]{aa}
\usepackage{natbib}

\usepackage{graphicx}
\usepackage{epsfig}
\usepackage{amssymb}

\newcommand{\bnabla}{ {\mbox{\boldmath$\nabla$}} }
\newcommand{\bbv}{ {\bf v} }

\newcommand{\bB}{{\bf B}}
\newcommand{ \ez }{ {\rm\bf e}_z}
\newcommand{ \ex }{ {\rm\bf e}_x}
\newcommand{ \ey }{ {\rm\bf e}_y}

\newcommand{ \dotp }{ {\mbox{\boldmath$\cdot$}} }
\newcommand{ \tz }{ \tilde{z}}
\newcommand{ \tts }{ \bar{s}}
\newcommand{ \sxi }{ {\mbox{\boldmath$\xi$}} }
\newcommand{ \seta }{ {\mbox{\boldmath$\eta$}} }

\title{An absorbing boundary formulation for the stratified, linearized, ideal MHD equations based on an unsplit, convolutional perfectly matched layer}
\titlerunning{Convolutional Perfectly Matched Layer}

\author{{Shravan M. Hanasoge}\inst{1,2} \and 
{Dimitri Komatitsch}\inst{3,4} \and
{Laurent Gizon}\inst{1}}

\authorrunning{Hanasoge et al.}

\institute{{Max-Planck-Institut f\"{u}r Sonnensystemforschung, 
Max Planck Stra{$\beta$}e 2, 37191 Katlenburg-Lindau, Germany}
\and{Department of Geosciences, Princeton University, Princeton, NJ 08544, USA}\and
{Universit\'e de Pau et des Pays de l'Adour, CNRS $\&$ INRIA Magique-3D,\\
Laboratoire de Mod\'elisation et d'Imagerie en G\'eosciences UMR 5212,\\
Avenue de l'Universit\'e, 64013 Pau Cedex, France}
\and{Institut universitaire de France, 103 boulevard
Saint-Michel, 75005 Paris, France}}


\date{Received 3 March 2010\\
Accepted 29 July 2010}
\abstract {
Perfectly matched layers are a very efficient way to absorb waves on the outer edges of media. We present a stable convolutional 
unsplit perfectly matched formulation designed for the linearized stratified Euler equations. The technique as applied to the
 Magneto-hydrodynamic (MHD) equations requires the use of a sponge, which, despite placing the perfectly matched
 status in question, is still highly efficient at absorbing outgoing waves. We study solutions
of the equations in the backdrop of models of linearized wave propagation in the Sun. We test the numerical stability of the schemes
by integrating the equations over a large number of wave periods.}

\begin{document}
\maketitle

\thanks{E-mail:hanasoge@mps.mpg.de}





\section{Introduction}
\label{intro}
The choice of appropriate boundary conditions in simulations remains a significant challenge in computational physics. Amidst the vast 
number of options that exist, a widely invoked construct is that of the absorbing boundary, one that is expected to relieve the computational domain
of outgoing waves or other structures without affecting the solution in the region of interest \citep[for a review, see e.g.,][]{colonius}. A number of different methods exist to 
accomplish this, such as an attenuating `sponge' \citep[e.g.,][]{lui,colonius}, characteristics-based boundary conditions \citep[e.g][]{thompson}, 
an artificially imposed supersonic outward directed advective flow \cite[e.g.,][]{lui}, perfectly matched layers \citep[PMLs;][]{Berenger1994}, etc. Unfortunately,
a number of these methods is afflicted with disadvantages, a primary cause of concern being that of absorption efficiency. For example, the characteristics-based
boundary conditions do not perform very well when the incident waves at the boundary are significantly inclined - and a fair degree of reflection is observed. The sponges
are better at absorbing the waves and are relatively easy to implement; however, the reflectivity can still be substantial.

The criterion of low reflectivity is perhaps best satisfied by the PML formulation, developed first by \citet{Berenger1994} as an absorbing
layer for the Maxwell equations. The central idea of this technique lies in performing an analytic continuation of the wave vector into the complex plane.
Wave vectors perpendicular to the boundary are forced to assume a decaying form in the complex plane, thereby dramatically reducing the amplitudes of the waves
in the boundary PML region. In the absence of discretization errors, the PML as set out by \citet{Berenger1994} is highly absorbent. After numerical 
discretization however, there are weak associated reflections. An important issue in the classical split formulation of \citet{Berenger1994} is that the absorption efficiency decreases 
rapidly at grazing incidence \citep[e.g.,][]{CoMo98a,WiRa00}.

There have been numerous advances in constructing stable analytic continuations of wave vectors in the boundary region. In particular, the stable convolutional
PMLs \citep[C-PML; e.g.,][]{roden2000,FeVi05,Komatitsch2007,DrGi07b}, contain a Butterworth filter inside the PML, thereby dramatically improving the absorption efficiency at grazing incidence.
This formalism was adopted to the study of waves in anisotropic geophysical media by \citet{Komatitsch2007}; the numerical stability of said
technique was studied by \citet{Komatitsch2007,Meza2008}. Extensions to the poro- and viscoelastic cases have been introduced by 
\citet{MaKoEz08} and \citet{MaKo09}, respectively. Our goal in this contribution is to develop a C-PML for the 3D linearized ideal MHD equations in stratified media.

Astrophysical media such as stellar interiors and atmospheres may be strongly stratified and highly magnetized. Simulating wave propagation in such environments is of interest because
in understanding the effects of stratification and magnetic fields on the oscillations, we may be able to better interpret observations and constrain certain properties of the object in question.
In particular, we focus on the Sun, a star that has been well studied and for which high-quality observations exist.
The interior of the Sun is opaque to electromagnetic waves -  consequently, 
only photons that arrive from very close to the surface (photosphere) of the Sun are visible. 
Helioseismology is the inference of the internal structure and dynamics
of the Sun by observing the surface manifestation of its acoustic pulsations \citep[e.g.,][]{dalsgaard02,gizon05,gizon09}. Armed with accurate interaction theories of waves and measurements 
of the acoustic field at the surface, one can attempt to determine the
sub-surface structure and dynamics of various solar features such as sunspots \citep[see e.g.,][]{gizon_etal_2009}, large-scale meridional flow, interior convective length scales, etc. In order to construct these interaction theories, 
it is important to simulate
and study small amplitude (linear) wave propagation in a solar-like medium - filled with scatterers like sunspots or mean flows. The time scales of wave propagation are typically much smaller than
the rate at which the scatterer itself evolves; thus one may invoke the assumption of time stationarity of the background medium. Numerical simulations of waves in solar-like media have been performed by numerous
groups \citep[e.g.,][]{werne04,hanasoge1,shelyag06,khomenko06,cameron07,parchevsky_method,Hanasoge_couvidat08, hanasoge_mag,cameron08}. Two groups in particular \citep{khomenko06, parchevsky_method} 
currently utilize the classical split PML formulation in order to solve the wave equations in stratified media. There exist drawbacks in these formulations. For example, \citet{khomenko06} have noted that there are long
term instabilities associated with their method. The technique discussed in \citet{parchevsky_method} involves the addition of a small arbitrary constant (see Eq. [21] and related discussion of their article)
that could possibly be acting as a sponge; it is unclear if their method is perfectly matched after the introduction of this constant. The modal power spectrum that \citet{parchevsky_method} display
in Figure 9 of their article shows substantial reflected power from their lower boundary, atypical of perfectly matched formulations.

The plan of this article is as follows: in Section 2, we recall the linearized ideal MHD equations and discuss the C-PML method as was previously developed for the seismic wave equations. The C-PML is then extended to the
stratified Euler/MHD equations in Section 3 and results from numerical tests are discussed. We summarize and conclude in Section 4. 

\section{The linearized ideal MHD equations}
As discussed in the introduction, we consider a system with a steady background, the wave equations are written as small perturbations around this state. The
assumption of small amplitude wave perturbations is consistent with
observations of solar and stellar oscillations \citep[e.g.,][]{jcd_notes, bogdan00}. In the equations that follow, all quantities with a subscript `0' denote the steady background properties while the unsubscripted
terms represent the fluctuations around the corresponding property. All bold faced terms represent vectors. 
\begin{eqnarray}
\partial_t\rho &=& -\bnabla\dotp(\rho_0 \bbv), \label{cont.1}\\
\partial_t(\rho_0 \bbv) &=& \bnabla\cdot \left[\frac{\bB\bB_0}{4\pi} + \frac{\bB_0\bB}{4\pi} -\left(p + \frac{\bB_0\cdot\bB}{4\pi}\right)\underbar{\bf{I}}\right] - \rho{g_0}\ez \nonumber\\ 
&+& {\bf S}(x,y,z,t), \label{mom.1}\\
\partial_t p &=& -c_0^2\rho_0\bnabla\dotp\bbv - \bbv\dotp\bnabla p_0,\label{press.1}\\
\partial_t \bB &=& \bnabla\cdot\left( \bB_0\bbv- \bbv\bB_0\right),\label{ind.1}\\
\bnabla\cdot\bB &=& 0. \label{div.1}
\end{eqnarray}
In order, these are the equations of continuity (Eq. [\ref{cont.1}]), momentum (Eq. [\ref{mom.1}]), adiabaticity (Eq. [\ref{press.1}]), induction (Eq. [\ref{ind.1}]), 
and divergence-free field (Eq. [\ref{div.1}]). The terms $\rho$ and $p$ denote density and pressure, $\bB = (B_x, B_y, B_z)$ is the magnetic field, $g_0$ is the 
gravity, $c_0$ the sound speed, and $\bbv = (v_x, v_y,v_z)$ is the velocity. A Cartesian coordinate system $(x,y,z)$
with unit vectors $({\bf e}_x, {\bf e}_y, {\bf e}_z)$ is adopted. The identity tensor is $\underbar{{\bf I}} = \ex\ex + \ey\ey + \ez\ez$. The momentum equations are forced by an 
arbitrary spatio-temporally smooth source function ${\bf S}$. All background terms $p_0, \rho_0, c, \bB_0$ are assumed to be heterogeneous, i.e., functions of $(x,y,z)$ and $g_0 = g_0(z)$. 
Note that we have assumed the background medium to be non-moving, i.e., that $\bbv_0 = 0$. The extension to the $\bbv_0 \ne 0$ situation requires no additional treatment at the boundary 
as long as the background velocity is zero in this region.

\subsection{PMLs and C-PMLs}
Consider a wave propagating in the $z$ direction, towards the upper boundary. For now, ignoring the fact that the medium is stratified, we have $v_z\sim A e^{i(k_z z -\omega t)}$, where 
$A$ is the amplitude of the wave, $k_z$ is the wavenumber along the vertical direction, $\omega$ the frequency, and $t$, time. The classical PML of \citet{Berenger1994} prescribes the
following transformation for $k_z$:
\begin{equation}
k_z \rightarrow {k_z}\left[{1 - \frac{d}{i\omega}}\right], \label{pml.classic}
\end{equation}
where $d = d(z)f(x,y,z_0) \ge 0$ is a damping parameter and $z=z_0$ is the vertical coordinate corresponding to the entrance of the layer. We include the term $f(x,y,z_0)$ to account for the 
possibility of strong wave speed variations in the horizontal (non-PML) directions at $z= z_0$ (i.e., at the entrance of the PML). 
Note that the damping term applies only to $k_z$; i.e. the wavenumbers $k_x$ and $k_y$ remain unchanged.
It has been shown that, cast in the form of equation~(\ref{pml.classic}), the PML is unstable in a number of situations (for $f(x,y,z_0)=1$), including in the presence of mean flows \citep[e.g.,][]{Hu2001, appelo2006}, and
in anisotropic media \citep[e.g.,][also personal communication, Khomenko 2009]{BeFaJo03,Komatitsch2007}. A version of the PML with a more general transformation was introduced by \citet[][]{roden2000}
for Maxwell equations and adapted to the seismic wave equation by \citet{FeVi05}, \citet{Komatitsch2007}, and \citet{DrGi07b}. They applied the following relation:
\begin{equation}
k_z \rightarrow {k_z}\left[{\kappa + \frac{d}{\alpha - i\omega}}\right]\label{wave.trans1},
\end{equation}
where $\alpha = \alpha(z) > 0$ and $\kappa = \kappa(z) \ge 1$ are additional parameters. In particular, the inclusion of $\alpha$, i.e. of a filter, helped improve the issue of damping waves with grazing incidence at 
the boundaries. We follow the formalism set out in \citet{Komatitsch2007} and derive the equations for the convolutional PML. The transformation of equation~(\ref{wave.trans1}) corresponds to a grid stretching of
the relevant coordinate:
\begin{equation}
\tilde{z} = \left(\kappa + \frac{d}{\alpha - i\omega}\right)z\label{grid.stretch1}.
\end{equation}
Derivatives along the vertical direction in equations~(\ref{cont.1}) through~(\ref{div.1}) are now computed in terms of the stretched coordinate. In other words, the vertical derivative of a generic variable $\psi(x,y,z,t)$
is transformed according to:
\begin{equation}
\partial_z\psi \rightarrow \partial_{\tz}\psi.
\end{equation}
The term $\partial_{\tz}\psi$ may be rewritten as:
\begin{eqnarray}
\partial_{\tz}\psi &=& \left[\frac{1}{\kappa} -  \frac{d/\kappa^2}{ (d/\kappa + \alpha)-i\omega }\right] \partial_z\psi. \label{wave.trans2}
\end{eqnarray}
The first term inside the parenthesis on the right-hand side of equation~(\ref{wave.trans2}) is easily obtained by dividing the derivative by $\kappa$. The second term is more complicated because 
it involves products in the temporal Fourier
space of two terms with frequency content; this results in a convolution in time. In order to compute these convolutions, \citet{roden2000,FeVi05,Komatitsch2007} use a recursion formula to update
the convolution at each time step. Instead of applying a recursive relation, here we use a differential equation to recover $\chi(x,y,z,\omega) =  \tts\partial_{z}\psi$ \citep[also done in the
auxiliary differential equation formulation of][]{GeZh10, MaKoGeBr10}, where
\begin{equation}
\tts(x,y,z,\omega) = - \frac{d/\kappa^2}{ (d/\kappa + \alpha)-i\omega }.
\end{equation}
 Let the time history of $\chi$ be such that $\chi(x,y,z,t \le 0) = 0$. In addition, we also require $\partial_z\psi = 0$ 
for all $t\le 0$. Then, the following differential equation for $\chi$
\begin{equation}
\partial_t \chi = -\frac{d}{\kappa^2}\partial_{z}\psi - \left(\frac{d}{\kappa} + \alpha\right) \chi \label{control.1},
\end{equation}
ensures that $\chi$ has the desired frequency response, i.e.,
\begin{equation}
\chi(x,y,z,\omega) =  - \frac{d/\kappa^2}{ (d/\kappa + \alpha)-i\omega } \partial_z\psi.
\end{equation}

The differential equations for the C-PML may now be written as:
\begin{eqnarray}
\partial_t\rho &=& -\bnabla_h\dotp(\rho_0 \bbv) - \rho_0\partial_{\tz}v_z - v_z\partial_{\tz} \rho_0,\label{mod.cont}\\
\partial_t \left(\rho_0\bbv\right) &=&   \bnabla_h\cdot \left[\frac{\bB\bB_0}{4\pi} + \frac{\bB_0\bB}{4\pi} -\left(p + \frac{\bB_0\cdot\bB}{4\pi}\right)\bf{I}\right] \nonumber\\
&+& \partial_{\tz}\left[\frac{B_z\bB_0}{4\pi} + \frac{B_{0z}\bB}{4\pi} -\left(p + \frac{\bB_0\cdot\bB}{4\pi}\right)\ez\right] \nonumber\\
&-& \rho \tilde{g}_0\ez  - \sigma\rho_0\bbv + {\bf S},\\
\partial_t p =&-&c_0^2\rho_0\bnabla_h\dotp\bbv - c^2\rho_0\partial_{\tz}v_z - \bbv\dotp\bnabla_h p_0 - v_z\partial_{\tz} p_0,\label{adiab.pml}\\
\partial_t \bB &=& \bnabla_h\cdot\left( \bB_0\bbv- \bbv\bB_0\right) + \partial_{\tz}\left(B_{0z}\bbv - v_z\bB_0\right),\label{mod.induction}
\end{eqnarray}
where $\bnabla_h = {\bf e}_x\partial_x +  {\bf e}_y\partial_y$ represents the derivatives in the non C-PML directions, and $\tilde{g}_0, \partial_{\tz}{p_0},\partial_{\tz}{\rho_0}$ are the 
modified background gravity, pressure and density gradients in
the absorption region respectively. A sponge-like decay term $-\sigma\rho_0\bbv$, where $\sigma= \sigma(x,y,z)$ is introduced in the momentum equation in order to stabilize the system. It
 is important to note that in the absorbing layer, the solution is non-physical and therefore discarded. Thus any 
modifications applied to the equations, in relation to issues such as altering the stratification or not enforcing $\bnabla\cdot\bB=0$, is justifiable as long as the solution in the interior 
domain is unaffected and the formulation is stable. The equations~(\ref{mod.cont}) through~(\ref{mod.induction}) in conjunction with equation~(\ref{control.1}) provide
the following prescription:

\begin{eqnarray}
\partial_t\rho &=& -\bnabla_h\dotp(\rho_0 \bbv) - \rho_0\left[\frac{1}{\kappa}\partial_{z}v_z + \Psi \right] - v_z\partial_{\tz}{\rho_0},\label{cont.2}\\
\partial_t\Psi &=& -\frac{d}{\kappa^2}\partial_{z}v_z - \left(\frac{d}{\kappa} + \alpha\right) \Psi,\label{memory.1}\\
\partial_{\tz}\rho_0 &=& \frac{\alpha/\kappa}{d/\kappa + \alpha} \partial_z\rho_0,\label{supp.rho}\\
\partial_t \left(\rho_0\bbv\right) &=&  \bnabla_h\cdot \left[\frac{\bB\bB_0}{4\pi} + \frac{\bB_0\bB}{4\pi} -\left(p + \frac{\bB_0\cdot\bB}{4\pi}\right)\underbar{\bf{I}}\right] \nonumber\\
&+& \frac{1}{\kappa}\partial_{z}\left[\frac{B_z\bB_0}{4\pi} + \frac{B_{0z}\bB}{4\pi} -\left(p + \frac{\bB_0\cdot\bB}{4\pi}\right)\ez\right]   \nonumber \\
 &&+ \sxi- \rho \tilde{g}_0\ez -\sigma\rho_0\bbv + {\bf S},\\
\partial_t\sxi &=& - \frac{d}{\kappa^2}\partial_{z}\left[\frac{B_z\bB_0}{4\pi} + \frac{B_{0z}\bB}{4\pi} -\left(p + \frac{\bB_0\cdot\bB}{4\pi}\right)\ez\right]  \nonumber\\
&&- \left(\frac{d}{\kappa} +  \alpha\right)\sxi, \label{memory.2}\\
\tilde{g}_0 &=&  \frac{\alpha/\kappa}{d/\kappa + \alpha} g_0,\label{supp.g}\\
\partial_t p &=&-c_0^2\rho_0\bnabla_h\dotp\bbv - c^2\rho_0\left[\frac{1}{\kappa}\partial_{z}v_z + \Psi \right]\nonumber\\
 &-& \bbv\dotp\bnabla_h p_0 - v_z\partial_{\tz} p_0,\\
\partial_{\tz}p_0 &=& \frac{\alpha/\kappa}{d/\kappa + \alpha} \partial_zp_0,\label{supp.p0}\\
\partial_t \bB &=& \bnabla_h\cdot\left( \bB_0\bbv- \bbv\bB_0\right) \nonumber\\
 &&+ \frac{1}{\kappa}\partial_{z}\left(B_{0z}\bbv - v_z\bB_0\right) + \seta,\\
\partial_t\seta &=& -\frac{d}{\kappa^2} \partial_z\left(B_{0z}\bbv - v_z\bB_0\right) -  \left(\frac{d}{\kappa} + \alpha\right)\seta.\label{memory.3}
\end{eqnarray}
Note that $\sxi,\seta$ are vector memory variables required for the calculation of the convolutional part of equation~(\ref{wave.trans2}) and that $\seta = (\eta_x, \eta_y, 0)$. 
These are additional variables are needed in order to solve the auxiliary differential equations~(\ref{memory.1}),~(\ref{memory.2}), and~(\ref{memory.3}).

Why is the stratification altered in the boundary region? Consider for example, the derivative term $\partial_{\tz} p_0$.
The differential equation~(\ref{control.1}) when applied to the steady $\partial_z p_0$ term leads to:
\begin{eqnarray}
\partial_{\tz}p_0 &= &\frac{1}{\kappa} \partial_zp_0 + \chi\\
\partial_t \chi &=& -\frac{d}{\kappa^2}\partial_z p_0 - \left(\frac{d}{\kappa} + \alpha\right) \chi\label{eqsec}.
\end{eqnarray}
Since we are dealing with a steady background medium, we drop the time derivative in equation~(\ref{eqsec}) and obtain
\begin{equation}
\partial_{\tz}p_0 = \frac{\alpha/\kappa}{d/\kappa + \alpha} \partial_zp_0.
\end{equation}  
Note that we can pursue a similar formalism for the gradient term $\partial_{\tz}\rho_0$. To maintain hydrostatic stability in the absorption layer, we alter gravity by prefixing it by the same term,
$\frac{\alpha/\kappa}{d/\kappa + \alpha}$. And hence the relations in equations~(\ref{supp.rho}),~(\ref{supp.g}), and~(\ref{supp.p0}). 

In the presence of magnetic field, three wave branches appear: magneto-sonic slow, fast, and Alfv\'{e}n (the distinctions between these modes apply only if the
magnetic and hydrodynamic pressures are not in equipartition). The physics of the propagation of these waves is intricate and inherently highly anisotropic \citep[e.g.,][]{goedbloed2004}. 
Alfv\'{e}n waves are transverse incompressible
oscillations whereas the slow and fast modes are compressive magnetically guided/influenced oscillations. 
The slow mode dispersion relation exhibits certain pathologies wherein for certain wavenumbers, the phase speed vanishes while the group speed becomes very small
 \citep[e.g., see pp. 195-214 of][]{goedbloed2004}. This strong anisotropy destabilizes the calculation in the entry regions of the boundary layer. Thus to offset the possibility of
this instability, we add the sponge term $-\sigma\rho_0\bbv$ to the momentum equations in regions of the magnetic field.  

The choice of the parameters $\kappa, d, \alpha$ is fairly central to creating an efficient absorption boundary formulation. Following \citet{Komatitsch2007}, we set $\alpha = \pi f_0$, where $f_0$ is a characteristic
frequency of the waves. The $\alpha$ decays linearly to zero over the absorption region, being maximum at the start of the layer and falling to zero at the boundary.
The damping function $d = f(x,y,z_0)~(z/L)^N$, is zero at the boundary layer entry point and attains its maximum value at the boundary. We find that for the Euler equations, the following choice for $f$ results in
efficient absorption \citep[e.g.,][]{CoTs01}:
\begin{equation}
 f(x,y,z_0) = -\frac{N+1}{2L} \bar{c} \log_{10} R_c, \label{absband}
\end{equation}
where $L$ is the length of the layer, $R_c$ is the amount of reflection tolerated, and $\bar{c}$ is the characteristic sound speed in the C-PML layer. Note that $\bar{c}$ is not a function of $z$. For the ideal MHD equations,
we use:
\begin{equation}
 f(x,y,z_0) = -\frac{N+1}{2L} c_w \log_{10} R_c, 
\end{equation} 
where, $c_w(x,y,z_0) = \sqrt{c_0(x,y,z_0)^2 + c_A(x,y,z_0)^2}$ is the characteristic propagation speed of the fastest waves in the absorption layer (set for a specific 
vertical location $z=z_0$) and $R_c, N, L$ retain the same meaning as for the Euler equations. Wave propagation speeds can vary substantially
from strongly magnetized regions to non-magnetic regions; for this reason, we allow $c_w$ to vary horizontally, i.e., $c_w = c_w(x,y,z_0)$. The sound speed is denoted by $c$ and
the Alfv\'{e}n speed by $c_A = ||\bB||/\sqrt{4\pi\rho_0}$.
The $\kappa$ term acts to damp evanescent waves \citep{roden2000,Ber02a}; although there are evanescent waves in our simulations, we find that instabilities are set into motion when $\kappa$ is allowed to vary from 
1 at the absorption layer entry to 8 at the boundary. For values of $\kappa$ less than this, the improvement in absorption efficiency is not very noticeable. We therefore set $\kappa =1$. The damping sponge is
$\sigma = -[(N+1)c_A/L]  (z/L)^N \log_{10} R_c$. The perfectly matched status of the MHD formulation comes into question with the introduction of the sponge
term. Note that because $\sigma \propto c_A$ the technique is perfectly matched for the stratified Euler equations, i.e. for the $c_A = 0$ case. 



\section{Numerical results}
\subsection{Methods}
The simulation code employed in these calculations has been discussed in various previous publications \citep{hanasoge_thesis, Hanasoge_couvidat08, hanasoge_mag}. The 
validation and verification of the numerical aspects of the code may also be found in these articles. In summary, to capture variations in the vertical direction, we apply sixth order, 
low dissipation and dispersion compact finite differences \citep{lele92}. A non-uniformly spaced grid is used because the eigenfunctions change much more gradually 
deeper within the interior; the vertical grid spacing varies from several hundred kilometers deeper within to tens of kilometers in the near-surface layers. Because the dissipation
of the scheme is low, spectral blocking occurs along the vertical directions (energy buildup at the highest wavenumbers). This is a direct consequence of the strong
vertical stratification (terms like $c, \rho_0, p_0$ vary strongly with $z$) and the non-uniform grid spacing; in order to avoid instabilities, we regularly de-alias the variables along the 
$z$ direction according to the technique described in \citet{dealias}. Zero Dirichlet upper and lower boundary conditions are enforced at the outer edge of the C-PMLs for all the variables.
A Fourier spectral method is applied in the horizontal directions in order to compute derivatives. 
Time stepping is achieved through the application of a second-order optimized Runge-Kutta scheme \citep{hu}. The horizontal boundaries are chosen to be 
periodic while the vertical boundaries are required to be absorbent. We study this case in detail below; there is no loss of generality however, since the method can be extended in order
to render the horizontal boundaries absorbent as well.

An inspection of equations~(\ref{cont.2}) to~(\ref{memory.3}) reveals that 6 memory variables are needed: for $v_z$, the three (vector) Lorentz 
force components, and two in the induction equation. The memory and computational overheads are relatively small: (1) the boundary layer works well with 8-10 grid points; thus the six memory variables  
need only be stored over a small number of points, and (2) the additional computations are limited to a small number of multiplications and additions, as required for
the evolution of the convolution equations~(\ref{memory.1}),~(\ref{memory.2}), and~(\ref{memory.3}).

\subsection{Waves in a non-magnetic stratified fluid}\label{stratified_nonmag}
\begin{figure}
\includegraphics[width=\linewidth]{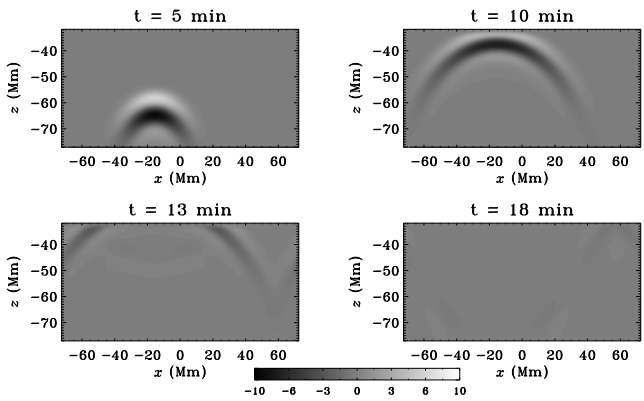}
\caption{Snapshots of the normalized wave velocity $\sqrt\rho_0 v_z$ at four time instants, non-dimensionalized to span the range $\pm 10$. The grey scale in all the panels is identical. The velocities at $t=18$ minutes are barely
discernible.}
\label{pmlq}
\end{figure}

\begin{figure}
\includegraphics[width=\linewidth]{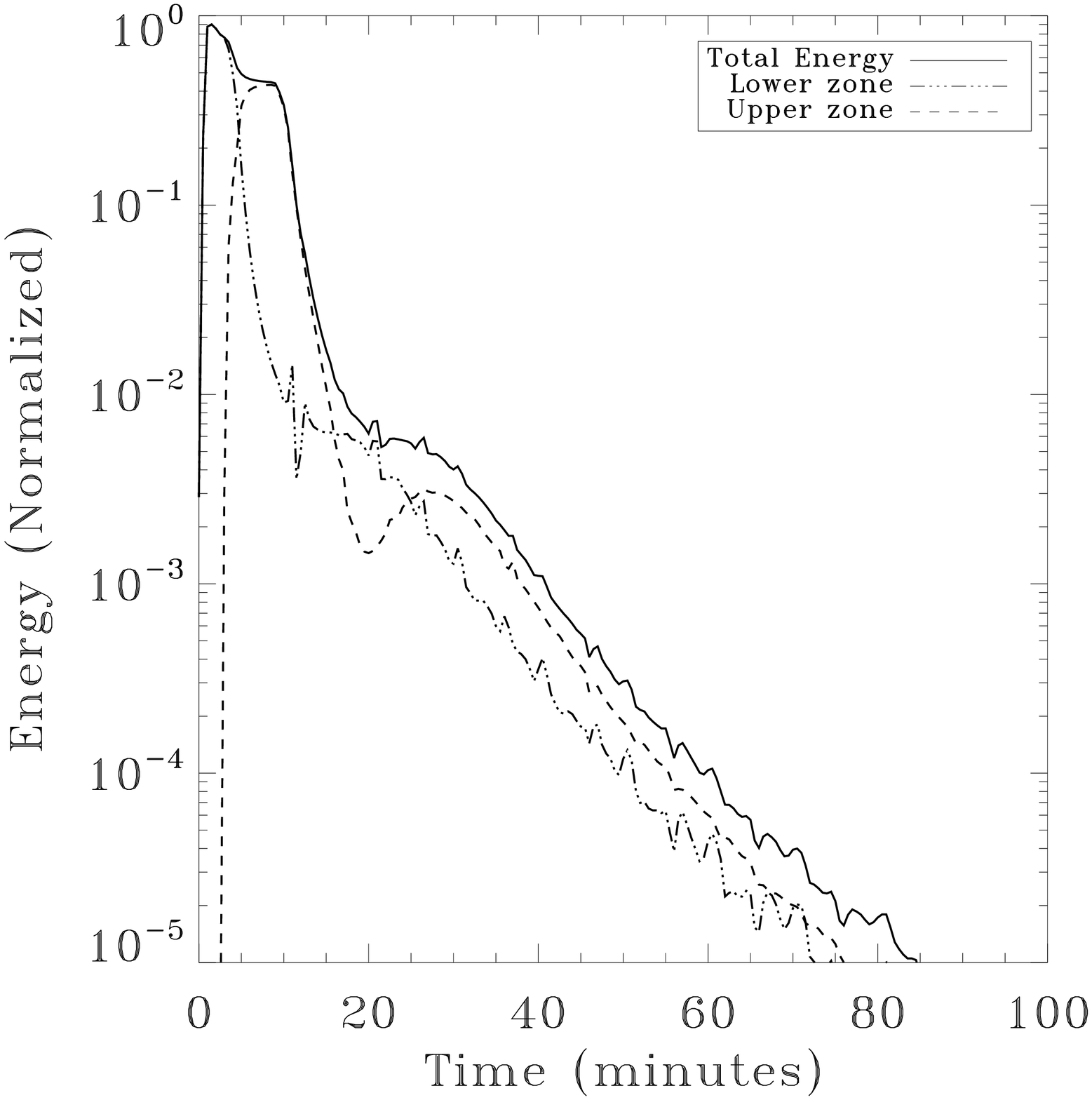}
\caption{Time evolution of the normalized modal energies as computed with equation~(\ref{energy.modes}). The initial drop in energy is due to the arrival of waves at the lower C-PML from the source.
{ The energy of these downward propagating waves is plotted as the dot-dash line.}
The second drop in the energy corresponds to the first arrivals at the upper C-PML { (displayed with dashed lines)}. Modes with large horizontal wave numbers and 
weakly reflected waves arrive gradually at the boundaries at later times, resulting in an extended energy decay. { Both upward and downward propagating wave packets are efficiently absorbed by the
CPMLs.}}
\label{energyq_time}
\end{figure}

We first demonstrate the ability of the C-PMLs at absorbing outgoing waves. The background medium is a stratified polytrope with index $m=2.15$; the vertical extent of the computational
domain is such that approximately 2.6 scale heights in density and 3.72 scale heights in pressure fill the domain (i.e. the pressure at the bottom is $e^{3.72} = 41.5$ times the value at the top). The 
stratification properties of the polytrope are described in appendix~\ref{only.polytrope}.
Waves are locally excited in a small
region located at a distance of 18 Mm above the bottom boundary (the full vertical length is 68 Mm).  For the C-PML parameters, we choose $N=2, R_c = 0.1\%, f_0 = 0.005 ~{\rm Hz}$ 
in equations~(\ref{eqsec}) and~(\ref{absband}) for these calculations.
Both the upper and lower C-PMLs are 10 grid points thick. In Figure~\ref{pmlq}, we show snapshots of the wavefronts at four instants of time. The
scaling in all the plots is identical. It is seen that the waves are almost completely absorbed by the upper boundary. In order to study the absorption more quantitatively, we
plot the following energy invariant (summed over the entire grid) as a function of time in Figure~\ref{energyq_time}, \citep[e.g.,][]{bogdan96}:
\begin{equation}
e = \frac{1}{2}\rho_0||\bbv||^2 +  \frac{p^2}{2\gamma p_0}.\label{energy.modes}
\end{equation}

In order to test the stability of the C-PMLs, we integrate the hydrodynamic wave equations for around 300 wave periods (peak frequency $\sim 6$ mHz; 12 hour time integration). In this case, 
the setup is more realistic; the computational domain now straddles the solar photosphere, a region where the density and pressure drop exponentially with height. The waves propagate in a domain that 
contains 21 scale heights in density, representing a contrast of 1.3 billion between the bottom and top. For details of the stratification, see appendix~\ref{full.stratification}.

We solve these equations in a computational domain of size $200 \times 200 \times 35$ Mm$^3$, (horizontal sides followed by vertical length) ($1 ~{\rm Mm} =10^6 {\rm m}$),
discretized using $256\times256\times300$ grid points. Vertically, the box extends from approximately 34 Mm below the solar photosphere to 1 Mm above.  Standing waves or normal modes 
are naturally created by the stratification of the medium; the modes are trapped between the surface and a specific depth below that depends on the frequency and wavenumber 
of the mode. A consequence of this trapping is that the mode has an evanescent tail that extends from the surface into the overlying atmosphere. We therefore place the upper 
computational boundary high enough above the surface so that the normal mode evanescent tails are largely unaffected by it.

 An additional aspect is that of the wave excitation mechanism: in the Sun, waves are created by the 
action of the strong near-photospheric turbulence \citep[e.g.,][]{stein00}. To mimic this process, we add a horizontally homogeneous vertically localized deterministic forcing 
function ($\bf S$ of the momentum equations). { This source term is only vertically directed, i.e., ${\bf S} = S(x,y,t)\delta(z-z_e)\ez$, where $z_e = - 50$ km. In order to generate $S$, 
we use a Gaussian random number generator to populate its Fourier domain and the resultant function is multiplied by the average frequency power spectrum of the 
Sun, modeled as a Gaussian with peak frequency at 3 mHz and a full width of 1 mHz. The inverse transform of this function is $S(x,y,t)$ \citep[e.g.,][]{hanasoge1, hanasoge_mag}.}

\begin{figure}
\includegraphics[width=\linewidth]{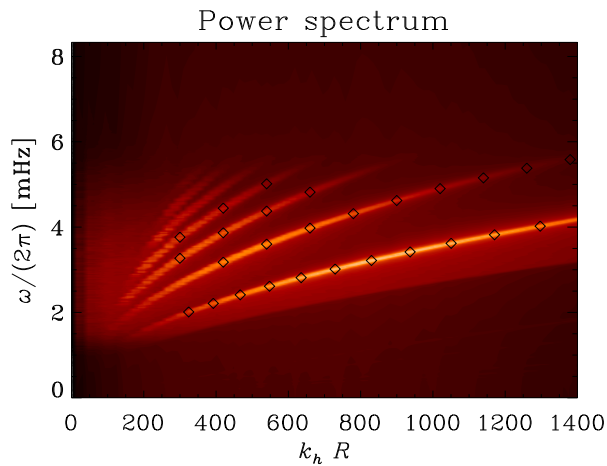}
\caption{Modal power spectrum of the vertical velocity component extracted at a height of $z = 200$ km above the photosphere, from a 12 hour long simulation with C-PMLs placed 
at the upper and lower boundaries { (plotted on a linear scale)}. The 
horizontal axis is the normalized wavenumber, the vertical is frequency, and regions
of high power are  the normal modes that appear in the calculation. The symbols overplotted on the power contours are the theoretically expected values of the resonant 
frequencies, computed using the MATLAB boundary value problem solver \texttt{bvp4c}.}
\label{powerq}
\end{figure}

The calculation is stable over the entire time integration. The wavenumber-frequency power spectrum of the vertical velocity component $v_z$ at a height of $z = 200$ km above
the photosphere is plotted in Figure~\ref{powerq}. The horizontal axis represents the non-dimensional horizontal wavenumber $k_h R_\odot$, where $R_\odot$ is the solar radius and $k_h$ is the horizontal 
 wavenumber, while the vertical axis is frequency in mHz.

\begin{figure}
\includegraphics[width=\linewidth]{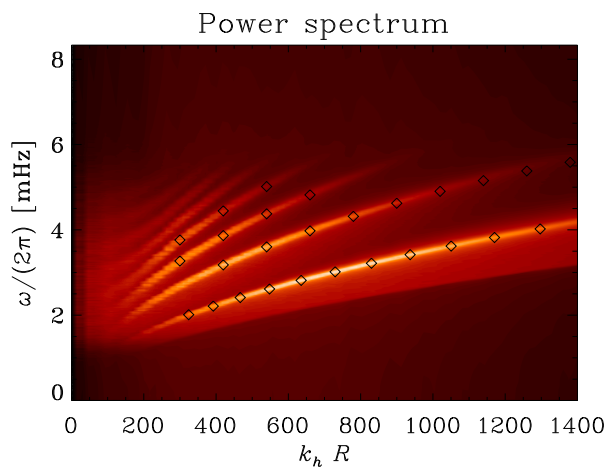}
\caption{Modal power spectrum extracted of the vertical velocity component extracted at a height of $z = 200$ km above the photosphere, 
from a 12 hour long simulation with the boundaries lined by sponges { (plotted on a linear scale)}. The weak reflections from the lower boundary cause the dispersion relation to
change in curvature at all points for which $\nu/(k_h R_\odot)$ exceeds a certain threshold. The symbols are the same as those in Figure~\ref{powerq}; note that there are differences between the C-PML
and sponge layered simulations at high frequencies. This is because these high-frequency waves are sensitive to the upper boundary.}
\label{spongepower}
\end{figure}

In order to study the impact of these absorbing layers, we perform simulations with C-PMLs placed adjacent to the upper and lower boundaries (Figure~\ref{powerq}) and damping sponges (Figure~\ref{spongepower}).
A demonstration of the efficacy of the C-PML is the absence of artifacts related to the lower boundary in the modal power spectrum in Figure~\ref{powerq}. Weak reflections
from the lower boundary cause the dispersion relation to change, manifested in the power spectrum as a flattening of the curvature of the ridges, as seen in Figure~\ref{spongepower}.

\subsection{Stratified MHD fluid}\label{stratified.mhd}
\begin{figure}
\includegraphics[width=\linewidth]{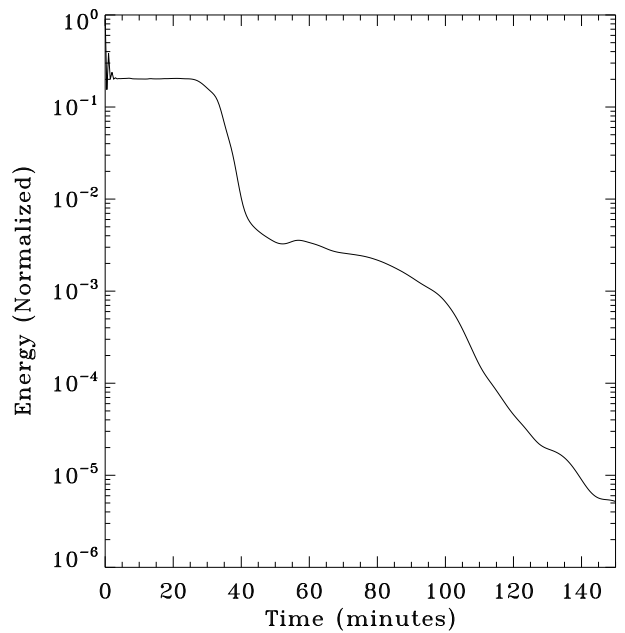}
\caption{Time evolution of the normalized MHD wave energy as computed using equations~(\ref{magenergy}), summed over the entire grid. 
Because of the strong dispersive nature of the waves and large variations in the propagation speeds between
the fast, slow, and Alfv\'{e}n waves, some modes take a long time to reach the boundaries, resulting in a relatively extended energy decay process compared to Figure~\ref{energyq_time}. 
The initial drop at $t\sim 20-40$ min corresponds to the first arrivals of waves 
at the upper and lower absorption layers. The secondary decline 
may be associated with the arrivals of the downward propagating dispersing slow and Alfv\'{e}n modes at the absorption layer adjacent to the lower boundary.}
\label{emag_time}
\end{figure}

We first perform a test akin to that of Figure~\ref{energyq_time}. A constant inclined magnetic field is embedded in the polytrope of appendix~\ref{only.polytrope}; the strength of the field is 
such that the Alfv\'{e}n speed is approximately four times the sound speed at the upper boundary. Waves are excited in the vicinity of the upper boundary by horizontally shaking the field lines. The fast
and Alfv\'{e}n waves reach the upper boundary first but the slow modes follow behind, which makes the demonstration of the absorptive properties of the boundary formulation less evident  
than in Figure~\ref{energyq_time}. Furthermore, constructing energy invariants for MHD waves in stratified media is not a trivial affair due to the fairly complicated Lorentz force terms. A 
number of authors have studied this problem \citep[e.g.,][]{Bray1974,Parker1979, Leroy1985} and have arrived
at differing versions of an invariant (personal communication, P. S. Cally 2009); we use the following approximate form \citep[kinetic + thermal + magnetic energies; e.g.,][]{Bray1974}:  
\begin{equation}
e = \frac{1}{2}\rho_0||\bbv||^2 +  \frac{p^2}{2\gamma p_0} + \frac{||\bB||^2}{8\pi}. \label{magenergy}
\end{equation}
 We plot the time history of the energy summed over the entire computational domain in Figure~\ref{emag_time}.

\begin{figure}
\vspace{0.5cm}
\includegraphics[width=0.9\linewidth]{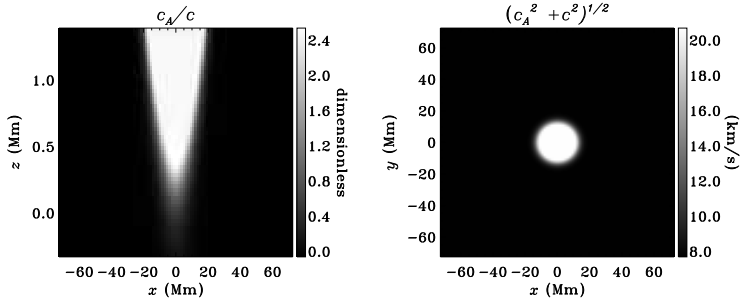}
\caption{The `sunspot' like magnetic field configuration used in the long-term integration numerical test. On the left panel we plot the ratio of the local Alfv\'{e}n to the sound speed as a function
of $x$ and $z$. The magnetic field becomes dynamically unimportant in the deeper layers because the background hydrostatic pressure increases with relative rapidity.
The right panel displays a horizontal cut (at the solar photosphere) through the fast mode speed ($\sqrt{c_A^2 + c_0^2}$) .
}
\label{magnetic.field}
\end{figure}

\begin{figure}
\includegraphics[width=\linewidth]{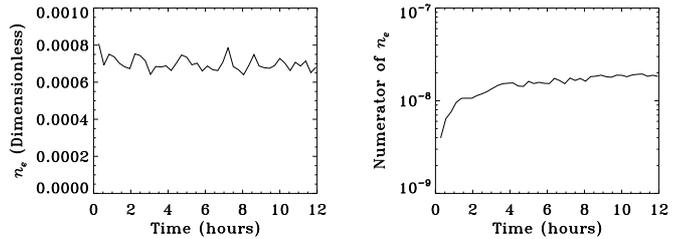}
\caption{Plotted on the left panel is the error in maintaining $||\bnabla\cdot\bB||=0$ as measured by $n_e$, defined in equation~(\ref{eq.ne}). On the right panel {is plotted
the $L_2$ norm of $\int_{z_1}^{z_2}dz~||\bnabla\cdot\bB||$ (numerator of Eq.~[\ref{eq.ne}]) as a function of time.}}
\label{divbhist}
\end{figure}

Our final test consists of studying the long-term stability of the method numerically. A magnetic flux tube \citep[e.g.,][]{moradi_hanasoge09}
is embedded in the stratified polytrope (appendix~\ref{full.stratification}); waves are excited in the non-magnetic regions and allowed to propagate through the 
flux tube, whose geometry is shown in Figure~\ref{magnetic.field}. The system is stable up to 12 hours of time integration. 
{ The modal power spectrum from this data looks almost identical to that in Figure~\ref{powerq} (with the exception of small frequency shifts) and is hence not shown.}
The maintenance of $\bnabla\cdot\bB=0$
numerically is an essential task of the scheme \citep[e.g.,][]{Toth2000}. Discretization errors are one of the primary causes for non-zero $\bnabla\cdot\bB=0$; additionally, in this case, the boundary
formulation in non $\bnabla\cdot\bB=0$ conserving. Consequently, it is important to test if the error grows with time and to quantify its level. In order to measure the normalized errors in 
maintaining $\bnabla\cdot\bB = 0$, we compute the following quantity:
\begin{eqnarray}
n_e &=& \frac{L_2\left[\int_{z_1}^{z_2}dz ||\bnabla\cdot\bB||\right]}{L_2\left[ ||\bB||\right]} \label{eq.ne},
\end{eqnarray}
where $z_1, z_2$ are the limits of the physically-relevant domain (i.e., the non-absorbing regions) and the following definition for the $L_2$ norm is applied:
\begin{equation}
L_2[f(x)] = \sqrt{\sum_i f(x_i)^2},
\end{equation}
where the sum is over all the grid points $x_i$ in the computational domain.
We display the time evolution of $n_e$ on the left panel of Figure~\ref{divbhist}. One can see that $n_e$ is nearly constant, perhaps showing a slight decrease with time.
On the right panel we show the time {history of the numerator of $n_e$ (Eq.~[\ref{eq.ne}])}. The gradual increase in the magnetic energy is due to constant
input wave excitation via the ${\bf S}$ term in the vertical momentum equation. The growth rates of the magnetic energy demonstrate the numerical stability of the system over the period of time integration. 


\section{Conclusions and Future Work}
We have developed a Convolutional Perfectly Matched Layer (C-PML) for the stratified linearized Euler equations and
a highly efficient absorption layer for the ideal MHD equations. 
This boundary formulation is quite useful for the calculation of wave propagation in astrophysical plasmas, where stratification and magnetic fields 
abound. The absorption layer for MHD waves is a slightly altered version of the convolutional formulation of \citep[e.g.,][]{roden2000}, requiring an extra sponge-like term 
in order to stabilize the system. While the boundary methods as discussed here are perfectly matched for the stratified Euler equations, the inclusion of the sponge term casts some 
doubt as to whether the same could be said of the MHD equations.  Simulations with the absorption layers for the stratified MHD and Euler equations were performed 
and found to be stable over a time length of 300 wave periods. With increasingly sophisticated boundary formulations, such as that of \citet{Hu2008}, who developed the 
PML for turbulent flows, we may in the near future be able to extend these techniques to fully non-linear MHD turbulence/dynamo calculations. 

\begin{acknowledgements}
S.M.H. is supported by the German Aerospace Center (DLR) through the
grant ``German Data Center for SDO". The study on sunspot modeling is
supported by the European Research Council under the European Community's
Seventh Framework Programme (FP7/2007-2013)/ERC grant agreement \#210949,
``Seismic Imaging of the Solar Interior", to PI L. Gizon. 
\end{acknowledgements}
\appendix
\section{Stratification}
\subsection{Only polytrope}\label{only.polytrope}
For the absorption tests in Sections~\ref{stratified_nonmag} and~\ref{stratified.mhd}, we use the following polytropic stratification prescription:
\begin{eqnarray}
p_0(z) &=& p_{\rm poly} \left(1 - \frac{z}{z_f}\right)^{m+1},\nonumber\\
\rho_0(z) &=& \rho_{\rm poly} \left(1 - \frac{z}{z_f}\right)^{m+1},\nonumber\\
g_0 &=& \frac{m+1}{z_f} \frac{p_{\rm poly}}{\rho_{\rm poly}},\label{polytrope.eq}\\
\Gamma_1 &=& 1 + \frac{1}{m},\nonumber\\
c_0(z) &=& \sqrt{\frac{\Gamma_1 p_{\rm poly}}{\rho_{\rm poly}}\left(1 - \frac{z}{z_f}\right)}.\nonumber\\
\end{eqnarray}
We set $p_{\rm poly} = 1.178 \times 10^5~{\rm dynes~ cm^{-2}}, \rho_{\rm poly} = 3.093 \times 10^{-7}~{\rm g~ cm^{-3}}, z_f = - 0.450~{\rm Mm},$ and $m=2.150$. Note also that $z$ is the height, i. e. the atmospheric
layers are described by $z>0$ and vice versa; $z=0$ is the fiducial surface. The polytrope is truncated at $z = -31.77$ Mm (upper boundary) and $z = -104.40$ Mm (lower boundary) 
respectively.

\subsection{Polytrope + isothermal layer}\label{full.stratification}
Waves propagating toward the surface in the Sun are reflected by a sharp fluctuation in the background density gradient \citep[e.g.,][]{jcd_notes}. This reflection zone is located at a 
height of $z_r\sim-0.050$ Mm below the surface. We mimic this by patching the polytropic stratification of section \ref{only.polytrope} with an overlying isothermal layer. This patching 
results in a fluctuation in the density gradient,
leading to an acoustic cut-off frequency of approximately 5.4 mHz, similar to the solar value. For $z < z_M$, we use the prescription of section~\ref{only.polytrope}. For $z \ge z_r$,
we use the following equations:
\begin{eqnarray}
p_0(z) &=& p_{\rm iso} \exp\left[\frac{z_r - z}{H}\right],\nonumber\\
\rho_0(z) &=& \rho_{\rm iso} \exp\left[\frac{z_r - z}{H}\right],\nonumber\\
\rho_{\rm iso} &=& \rho_{\rm poly} \left(1 - \frac{z_r}{z_f}\right)^{m},\\
p_{\rm iso} &=& p_{\rm poly} \left(1 - \frac{z_r}{z_f}\right)^{m+1},\nonumber\\
H &=& \frac{p_{\rm iso}}{g_0 \rho_{\rm iso}}.\nonumber
\end{eqnarray}
The relations for $\rho_{\rm iso}$ and $p_{\rm iso}$ arise from the requirement of continuity of the pressure and density at the matching point between the polytropic and isothermal
layers. The relation for $H$ is a consequence of enforcing hydrostatic balance. This model is truncated at $z = -34$ Mm (lower boundary) and $z=+1$ Mm (upper boundary).

\bibliographystyle{aa}
\bibliography{pml}

\begin{thebibliography}{48}
\expandafter\ifx\csname natexlab\endcsname\relax\def\natexlab#1{#1}\fi

\bibitem[{{Appel\"{o}} {et~al.}(2006){Appel\"{o}}, {Hagstrom}, \&
  {Kreiss}}]{appelo2006}
{Appel\"{o}}, D., {Hagstrom}, T., \& {Kreiss}, G. 2006, SIAM Journal of Applied
  Mathematics, 67, 1

\bibitem[{B\'ecache {et~al.}(2003)B\'ecache, Fauqueux, \& Joly}]{BeFaJo03}
B\'ecache, E., Fauqueux, S., \& Joly, P. 2003, Journal of Computational
  Physics, 188, 399

\bibitem[{B\'{e}renger(1994)}]{Berenger1994}
B\'{e}renger, J.-P. 1994, Journal of Computational Physics, 114, 185

\bibitem[{B\'erenger(2002)}]{Ber02a}
B\'erenger, J.~P. 2002, IEEE Microwave and Wireless Components Letters, 12, 218

\bibitem[{{Bogdan}(2000)}]{bogdan00}
{Bogdan}, T.~J. 2000, \solphys, 192, 373

\bibitem[{{Bogdan} {et~al.}(1996){Bogdan}, {Hindman}, {Cally}, \&
  {Charbonneau}}]{bogdan96}
{Bogdan}, T.~J., {Hindman}, B.~W., {Cally}, P.~S., \& {Charbonneau}, P. 1996,
  \apj, 465, 406

\bibitem[{{Bray} \& {Loughhead}(1974)}]{Bray1974}
{Bray}, R.~J. \& {Loughhead}, R.~E. 1974, {The solar chromosphere} (Chapman and
  Hall)

\bibitem[{{Cameron} {et~al.}(2007){Cameron}, {Gizon}, \&
  {Daiffallah}}]{cameron07}
{Cameron}, R., {Gizon}, L., \& {Daiffallah}, K. 2007, Astronomische
  Nachrichten, 328, 313

\bibitem[{{Cameron} {et~al.}(2008){Cameron}, {Gizon}, \& {Duvall}}]{cameron08}
{Cameron}, R., {Gizon}, L., \& {Duvall}, Jr., T.~L. 2008, \solphys, 251, 291

\bibitem[{{Christensen--Dalsgaard}(2003)}]{jcd_notes}
{Christensen--Dalsgaard}, J. 2003, Lecture Notes on Stellar Oscillations, 5th
  edn.

\bibitem[{{Christensen-Dalsgaard}(2002)}]{dalsgaard02}
{Christensen-Dalsgaard}, J. 2002, Reviews of Modern Physics, 74, 1073

\bibitem[{Collino \& Monk(1998)}]{CoMo98a}
Collino, F. \& Monk, P. 1998, Computational Methods in Applied Mechanical
  Engineering, 164, 157

\bibitem[{Collino \& Tsogka(2001)}]{CoTs01}
Collino, F. \& Tsogka, C. 2001, Geophysics, 66, 294

\bibitem[{Colonius(2004)}]{colonius}
Colonius, T. 2004, Annual Review of Fluid Mechanics, 36, 315

\bibitem[{Drossaert \& Giannopoulos(2007)}]{DrGi07b}
Drossaert, F.~H. \& Giannopoulos, A. 2007, Wave Motion, 44, 593

\bibitem[{Festa \& Vilotte(2005)}]{FeVi05}
Festa, G. \& Vilotte, J.~P. 2005, Geophysical Journal International, 161, 789

\bibitem[{Gedney \& Zhao(2010)}]{GeZh10}
Gedney, S.~D. \& Zhao, B. 2010, IEEE Transactions on Antennas and Propagation,
  58, 838

\bibitem[{{Gizon} \& {Birch}(2005)}]{gizon05}
{Gizon}, L. \& {Birch}, A.~C. 2005, Living Reviews in Solar Physics, 2, 6

\bibitem[{{Gizon} {et~al.}(2010){Gizon}, {Birch}, \& {Spruit}}]{gizon09}
{Gizon}, L., {Birch}, A.~C., \& {Spruit}, H.~C. 2010, Annual Review of
  Astronomy and Astrophysics

\bibitem[{{Gizon} {et~al.}(2009){Gizon}, {Schunker}, {Baldner}, {Basu},
  {Birch}, {Bogart}, {Braun}, {Cameron}, {Duvall}, {Hanasoge}, {Jackiewicz},
  {Roth}, {Stahn}, {Thompson}, \& {Zharkov}}]{gizon_etal_2009}
{Gizon}, L., {Schunker}, H., {Baldner}, C.~S., {et~al.} 2009, Space Science
  Reviews, 144, 249

\bibitem[{{Goedbloed} \& {Poedts}(2004)}]{goedbloed2004}
{Goedbloed}, J.~P.~H. \& {Poedts}, S. 2004, {Principles of
  Magnetohydrodynamics} (Cambridge University Press)

\bibitem[{{Hanasoge}(2007)}]{hanasoge_thesis}
{Hanasoge}, S.~M. 2007, PhD thesis, Stanford University, USA

\bibitem[{{Hanasoge}(2008)}]{hanasoge_mag}
{Hanasoge}, S.~M. 2008, \apj, 680, 1457

\bibitem[{{Hanasoge} {et~al.}(2008){Hanasoge}, {Couvidat}, {Rajaguru}, \&
  {Birch}}]{Hanasoge_couvidat08}
{Hanasoge}, S.~M., {Couvidat}, S., {Rajaguru}, S.~P., \& {Birch}, A.~C. 2008,
  \mnras, 391, 1931

\bibitem[{{Hanasoge} \& {Duvall}(2007)}]{dealias}
{Hanasoge}, S.~M. \& {Duvall}, Jr., T.~L. 2007, Astronomische Nachrichten, 328,
  319

\bibitem[{{Hanasoge} {et~al.}(2006){Hanasoge}, {Larsen}, {Duvall}, {DeRosa},
  {Hurlburt}, {Schou}, {Roth}, {Christensen-Dalsgaard}, \& {Lele}}]{hanasoge1}
{Hanasoge}, S.~M., {Larsen}, R.~M., {Duvall}, Jr., T.~L., {et~al.} 2006, \apj,
  648, 1268

\bibitem[{Hu(2001)}]{Hu2001}
Hu, F.~Q. 2001, Journal of Computational Physics, 173, 455

\bibitem[{{Hu} {et~al.}(1996){Hu}, Hussaini, \& Manthey}]{hu}
{Hu}, F.~Q., Hussaini, M.~Y., \& Manthey, J.~L. 1996, Journal of Computational
  Physics, 124, 177

\bibitem[{Hu {et~al.}(2008)Hu, Li, \& Lin}]{Hu2008}
Hu, F.~Q., Li, X., \& Lin, D. 2008, Journal of Computational Physics, 227, 4398

\bibitem[{{Khomenko} \& {Collados}(2006)}]{khomenko06}
{Khomenko}, E. \& {Collados}, M. 2006, \apj, 653, 739

\bibitem[{Komatitsch \& Martin(2007)}]{Komatitsch2007}
Komatitsch, D. \& Martin, R. 2007, Geophysics, 72, 155

\bibitem[{Lele(1992)}]{lele92}
Lele, S.~K. 1992, Journal of Computational Physics, 103, 16

\bibitem[{{Leroy}(1985)}]{Leroy1985}
{Leroy}, B. 1985, Geophysical and Astrophysical Fluid Dynamics, 32, 123

\bibitem[{{Lui}(2003)}]{lui}
{Lui}, C. 2003, PhD thesis, Stanford University, USA

\bibitem[{Martin \& Komatitsch(2009)}]{MaKo09}
Martin, R. \& Komatitsch, D. 2009, \gji, 179, 333

\bibitem[{Martin {et~al.}(2008)Martin, Komatitsch, \& Ezziani}]{MaKoEz08}
Martin, R., Komatitsch, D., \& Ezziani, A. 2008, Geophysics, 73, T51

\bibitem[{Martin {et~al.}(2010)Martin, Komatitsch, Gedney, \&
  Bruthiaux}]{MaKoGeBr10}
Martin, R., Komatitsch, D., Gedney, S.~D., \& Bruthiaux, E. 2010, Computer
  Modeling in Engineering and Sciences, 56, 17

\bibitem[{Meza-Fajardo \& Papageorgiou(2008)}]{Meza2008}
Meza-Fajardo, K.~C. \& Papageorgiou, A.~S. 2008, Bulletin of the Seismological
  Society of America, 98, 1811

\bibitem[{{Moradi} {et~al.}(2009){Moradi}, {Hanasoge}, \&
  {Cally}}]{moradi_hanasoge09}
{Moradi}, H., {Hanasoge}, S.~M., \& {Cally}, P.~S. 2009, \apjl, 690, L72

\bibitem[{{Parchevsky} \& {Kosovichev}(2007)}]{parchevsky_method}
{Parchevsky}, K.~V. \& {Kosovichev}, A.~G. 2007, Astrophysical Journal, 666,
  547

\bibitem[{{Parker}(1979)}]{Parker1979}
{Parker}, E.~N. 1979, {Cosmical magnetic fields: Their origin and their
  activity} (Oxford, Clarendon Press; New York, Oxford University Press, 858
  p.)

\bibitem[{{Roden} \& {Gedney}(2000)}]{roden2000}
{Roden}, J.~A. \& {Gedney}, S.~D. 2000, Microwave and Optical Technology
  Letters, 27, 334

\bibitem[{{Shelyag} {et~al.}(2006){Shelyag}, {Erd{\'e}lyi}, \&
  {Thompson}}]{shelyag06}
{Shelyag}, S., {Erd{\'e}lyi}, R., \& {Thompson}, M.~J. 2006, \apj, 651, 576

\bibitem[{{Stein} \& {Nordlund}(2000)}]{stein00}
{Stein}, R.~F. \& {Nordlund}, {\AA}. 2000, \solphys, 192, 91

\bibitem[{{Thompson}(1990)}]{thompson}
{Thompson}, K.~W. 1990, Journal of Computational Physics, 89, 439

\bibitem[{T{\'o}th(2000)}]{Toth2000}
T{\'o}th, G. 2000, Journal of Computational Physics, 161, 605

\bibitem[{{Werne} {et~al.}(2004){Werne}, {Birch}, \& {Julien}}]{werne04}
{Werne}, J., {Birch}, A., \& {Julien}, K. 2004, in ESA Special Publication,
  Vol. 559, SOHO 14 Helio- and Asteroseismology: Towards a Golden Future, ed.
  D.~{Danesy}, 172

\bibitem[{Winton \& Rappaport(2000)}]{WiRa00}
Winton, S.~C. \& Rappaport, C.~M. 2000, IEEE transactions on Antennas and
  Propagation, 48, 1055

\end{thebibliography}
\end{document}